\documentclass[twocolumn,prl,preprintnumbers,amsmath,amssymb,superscriptaddress,floatfix]{revtex4-2}

\usepackage{graphicx}
\usepackage{dcolumn}
\usepackage{bm}
\usepackage{color} 

\usepackage{amsmath,amssymb}
\def\lesssim{\ \raise.3ex\hbox{$<$}\kern-0.8em\lower.7ex\hbox{$\sim$}\ }
\def\gesim{\ \raise.3ex\hbox{$>$}\kern-0.8em\lower.7ex\hbox{$\sim$}\ }

\newcommand{\redd}[1]{{#1}}

\newcommand{\red}[1]{{#1}}

\newcommand \beq{\begin{eqnarray}}
\newcommand \eeq{\end{eqnarray}}
\begin{document}
\preprint{RIKEN-iTHEMS-Report-23}
\title{Exploring $^3P_0$ Superfluid in Dilute Spin-Polarized Neutron Matter}
\author{Hiroyuki Tajima}
\affiliation{Department of Physics, Graduate School of Science, The University of Tokyo, Tokyo 113-0033, Japan}
\affiliation{RIKEN Nishina Center, Wako 351-0198, Japan}

\author{Hiroshi Funaki}
\affiliation{%
Kavli Institute for Theoretical Sciences, University of Chinese Academy of Sciences, Beijing, 100190, China.
}%

\author{Yuta Sekino}
\affiliation{RIKEN Cluster for Pioneering Research (CPR), Astrophysical Big Bang Laboratory (ABBL), Wako, Saitama, 351-0198 Japan}
\affiliation{Interdisciplinary Theoretical and Mathematical Sciences Program (iTHEMS), RIKEN, Wako, Saitama 351-0198, Japan}

\author{Nobutoshi Yasutake}
\affiliation{Department of Physics, Chiba Institute of Technology (CIT), 2-1-1 Shibazono, Narashino, Chiba 275-0023, Japan}
\affiliation{%
Advanced Science Research Center, Japan Atomic Energy Agency, Tokai, 319-1195, Japan
}%

\author{Mamoru Matsuo}
\affiliation{%
Kavli Institute for Theoretical Sciences, University of Chinese Academy of Sciences, Beijing, 100190, China.
}%
\affiliation{%
Advanced Science Research Center, Japan Atomic Energy Agency, Tokai, 319-1195, Japan
}%
\affiliation{%
CAS Center for Excellence in Topological Quantum Computation, University of Chinese Academy of Sciences, Beijing 100190, China
}%
\affiliation{%
RIKEN Center for Emergent Matter Science (CEMS), Wako, Saitama 351-0198, Japan
}%

\begin{abstract}
We explore the theoretical possibility of $^3P_0$ neutron superfluid in dilute spin-polarized neutron matter, which may be relevant to the crust region of a magnetized neutron star.
In such a dilute regime where the neutron Fermi energy is less than 1~MeV, the $^1S_0$ neutron superfluid can be suppressed by a strong magnetic field of the compact star.
In the low-energy limit relevant for dilute neutron matter, the $^3P_0$ interaction is stronger than the $^3P_2$ one which is believed to induce the triplet superfluid in the core.
We present the ground-state phase diagram of dilute neutron matter with respect to the magnetic field and numerically estimate the critical temperature of the $^3P_0$ neutron superfluid, which is found to exceed $10^7$~K.
\end{abstract}
\maketitle

\noindent{\it Introduction}.---
A recent progress of neutron-star observations gives us an important opportunity to examine the exotic state of matter such as nucleon superfluidity.
The cooling process of neutron stars and pulsar glitch phenomena have been studied in connection with the neutron superfluidity~\cite{pines1985superfluidity}.
In the neutron-star crust region with the subnuclear density,
the $^1S_0$ neutron superfluid has been discussed extensively~\cite{RevModPhys.75.607} \red{(see also recent review~\cite{sedrakian2019superfluidity})}.
On the other hand, in the core region of neutron stars where the nucleon density is close to the normal nuclear density $\rho_0=0.16$~fm$^{-3}$,
the $^3P_2$ neutron superfluid is expected to occur based on the nucleon-nucleon phase shift at relevant energies with respect to the neutron Fermi energy therein~\cite{takatsuka1993superfluidity}.
Based on the recent rapid progress of multi-messenger astronomy observations~\cite{meszaros2019multi}, such nuclear many-body states can be further studied in  
various astrophysical environments \red{in the future, as the observed rapid cooling of Cassiopeia A implies the existence of $^3P_2$ neutron superfluid~\cite{PhysRevLett.106.081101}.}

\begin{figure}[t]
    \centering
    \includegraphics[width=8.6cm]{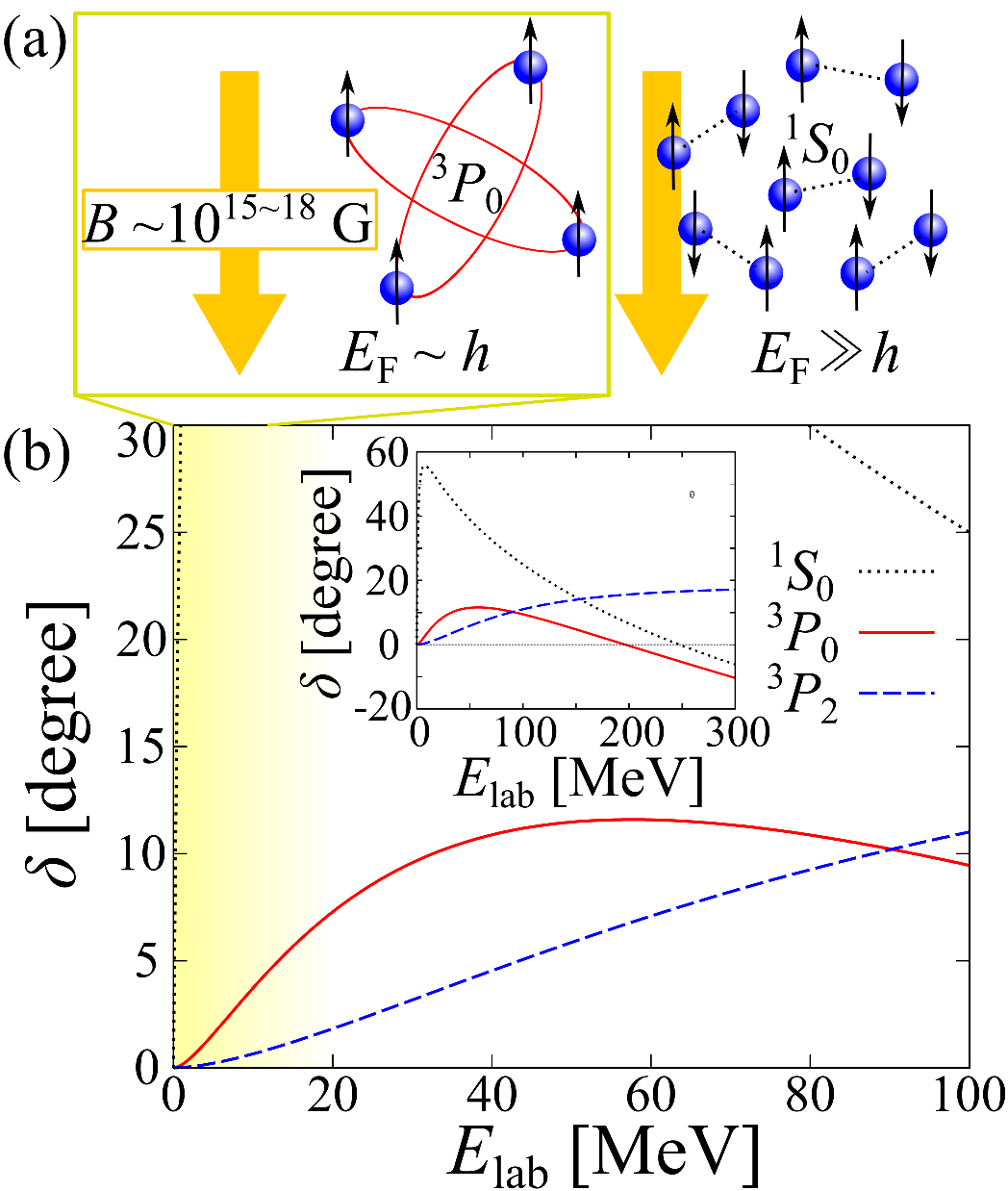}
    \caption{
    (a) Schematics of $^3P_0$ neutron superfluid in dilute neutron matter with a large spin polarization due to the magnetic field $B$ where the Zeeman shift $h$ is comparable with the Fermi energy $E_{\rm F}$.
    At larger densities where $E_{\rm F}\gg h$, the $^1S_0$ neutron superfluid without the spin polarization can appear. 
    (b) Low-energy nucleon-nucleon phase shift of the isovector channel from the Nijmegen partial wave analysis (NPWA)~\cite{PhysRevC.48.792} where $E_{\rm lab}$ is the laboratory kinetic energy. The inset shows the phase shift with the relatively high-energy regime ($\sim 300$~MeV).
    In the dilute region with $E_{\rm F}\sim h$ (shaded area), $^3P_0$ scattering phase shift can be a relevant channel for spin-polarized neutron matter. }
    \label{fig:1}
\end{figure}

The strong magnetic field in magnetars may lead to non-trivial effects not yet to be revealed (e.g., appearances of spin-$3/2$ $\Delta$ baryons~\cite{PhysRevC.106.035801} and superheavy nuclei~\cite{sekizawa2023possible}).
Magnetars may involve a strong magnetic field $B\sim 10^{15-18}$~G as studied in recent works, e,g., \red{Refs.~\cite{PhysRevC.99.055811,PhysRevC.107.045806,10.1093/mnras/stad1773}.}
\red{In particular, the deformation of the magnetar observed via X-ray spectra indicates the existence of an extreme toroidal magnetic field $B_{\rm t}$~\cite{PhysRevLett.112.171102}, which is stronger than the dipole one $B_{\rm d}$ estimated from the spin-down luminosity.}
The resulting Zeeman shift $h=|\gamma_{\rm n} B|/2\lesssim 10$~MeV with the neutron gyromagnetic ratio $\gamma_{\rm n}=-1.2\times 10^{-17}$~MeV/G
 is small compared to the neutron Fermi energy $E_{\rm F}$ around $\rho=\rho_0$ where $E_{\rm F}=\frac{(3\pi^2\rho)^{2/3}}{2M}\simeq 60$~MeV (where $M=939$~MeV is the neutron mass)~\cite{sedrakian2019superfluidity}. 
 This fact evinces that the $^1S_0$ neutron superfluid without the spin polarization should be dominant at subnuclear densities.
 However, it is not necessarily true in the dilute region where neutrons just started to drip from neutron-rich nuclei in the inner crust~\cite{PhysRevC.80.065801}.
At a smaller density $\rho=10^{-3}\rho_0$ (near the neutron drip density), one obtains $E_{\rm F}\simeq 0.6$~MeV, which can be comparable with the Zeeman shift $h$. 
As a result, $^1S_0$ pairing gap $\Delta_{^1S_0}\ll E_{\rm F}$ can be strongly suppressed by $h$~\cite{PhysRevC.93.015802}.

In such a dilute region, the low-energy nucleon-nucleon phase shift shown in Fig.~\ref{fig:1} is important.
One can see that $^1S_0$ channel is dominant at low energies ($E_{\rm lab}\lesssim 150$~MeV where $E_{\rm lab}$ is the laboratory kinetic energy).
However, suppose that dilute neutron matter is polarized due to the strong magnetic field,
the dominant attractive interactions turn to be the triplet $P$-wave channels, that is, $^3P_0$ and $^3P_2$.
Interestingly, the $^3P_0$ channel can be a leading contribution near the drip density \red{($E_{\rm lab}\lesssim 100$~MeV as shown in Fig.~\ref{fig:1})} 
in contrast to the core region of neutron stars ($E_{\rm lab}\gesim 150$~MeV corresponding to $E_{\rm F}\gg h$) where the $^3P_2$ channel is relevant~\red{\cite{PhysRevC.104.045803}}.

In this work, 
we theoretically explore the possibility of $^3P_0$ neutron superfluid in dilute spin-polarized neutron matter, which may be relevant to the crust region near the neutron drip line under the strong magnetic field.
First, we consider the possible ground-state phase diagram in terms of the magnetic field and the neutron density.
Because we are interested in the dilute regime where the low-energy universality is relevant,
we utilize the recent theoretical results of strongly interacting Fermi gases near the unitary limit~\cite{BCS-BEC,STRINATI20181,OHASHI2020103739} to characterize $^1S_0$ superfluid properties with largely negative scattering length $a=-18.5$~fm.
In such a way, we identify the possible spin-polarized regime at zero temperature.
Moreover, we develop the mean-field framework of $^3P_0$ neutron superfluid. 
The separable interaction is employed to reproduce the  $^3P_0$ scattering amplitude.
Finally, we predict the critical temperature of $^3P_0$ superfluid as a function of a spin-polarized neutron density.

Hereafter, we use the units of $\hbar=k_{\rm B}=c=1$ and the system volume is taken to be unity for convenience.

\begin{figure}[t]
    \centering
    \includegraphics[width=8.6cm]{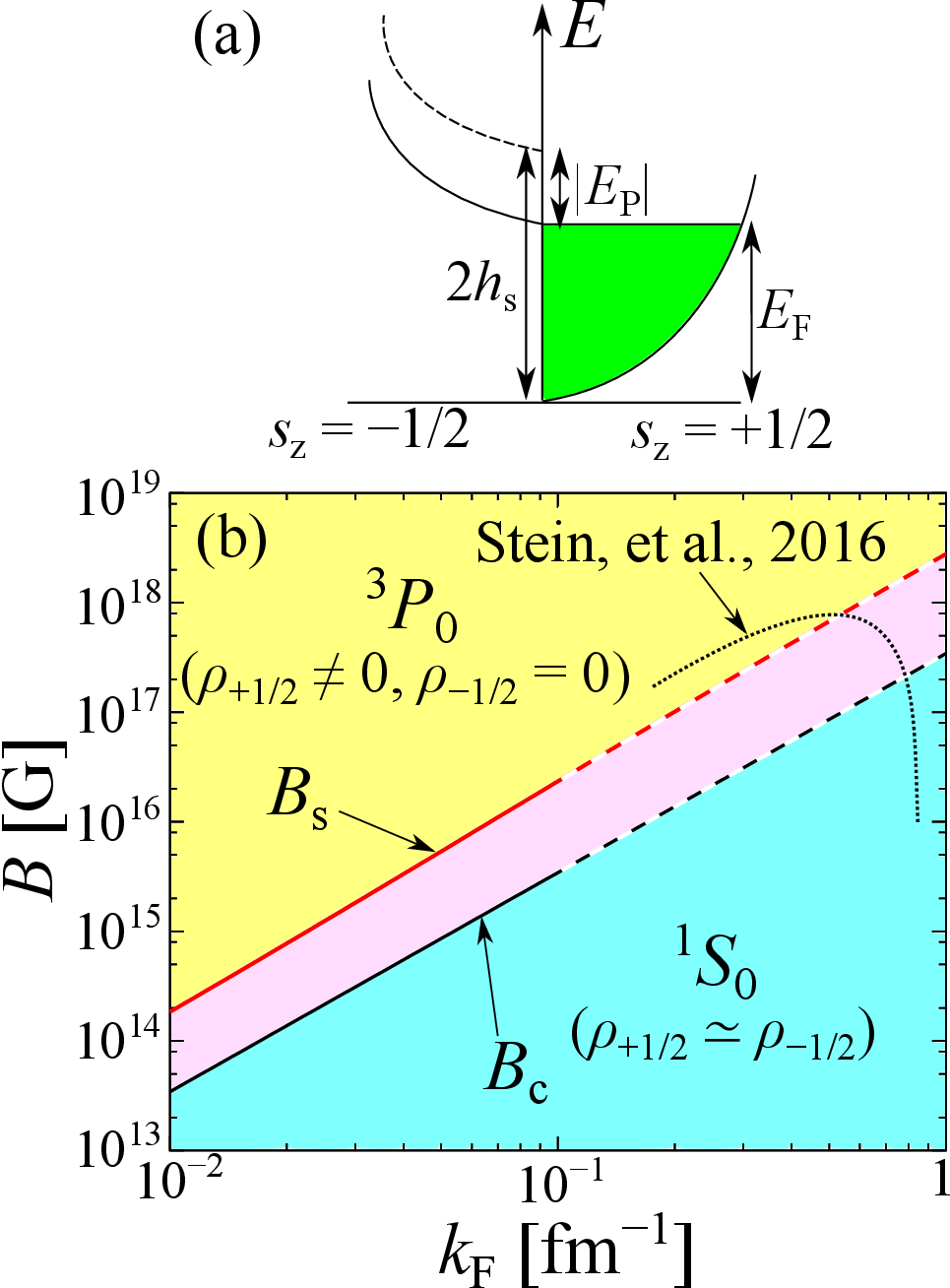}
    \caption{(a) Energy diagram of spin-$1/2$ neutrons (left: $s_z=-1/2$, right: $s_z=+1/2$) at the saturation Zeeman shift $h=h_{\rm s}$.
    While the $s_z=-1/2$ neutrons exhibit the polaron energy shift $E_{\rm P}$ with respect to the bare dispersion (dashed line),
    $s_z=+1/2$ neutrons form the Fermi sea with the Fermi energy $E_{\rm F}$.
    Below $h=h_{\rm s}$, the $s_{z}=-1/2$ state starts to be occupied.
    (b) Schematic ground-state phase diagram of dilute neutron matter under the strong magnetic field $B$. $k_{\rm F}$ is the Fermi momentum. $B_{\rm s}$ denotes the saturation magnetic field beyond which neutron matter is fully polarized.
    $B_{\rm c}$ is the critical magnetic field for the $^1S_0$ superfluid. 
    For comparison, the unpaired magnetic field in the mean-field calculation with the multi-rank separable interaction at $T=0.05$ MeV (Stein et al., 2016~\cite{PhysRevC.93.015802}) is also plotted (dotted curve).
    \redd{In the high-density region ($k_{\rm F}\gesim 0.1$ fm$^{-1}$) where the dashed lines are plotted, our results are not quantitatively valid due to the finite-range effect.} 
    }
    \label{fig:2}
\end{figure}

\noindent{\it Ground-state phase diagram}.---
First, we qualitatively examine the possible ground-state phase diagram in dilute neutron matter under the strong magnetic field.
To this end, the information of strongly interacting Fermi gases near the unitary limit is useful~\cite{BCS-BEC,STRINATI20181,OHASHI2020103739}.
In the dilute system with the negligible finite-range effect~\cite{PhysRevA.97.013601}, the relevant energy scale for the $^1S_0$ neutron superfluid is given by $1/(Ma^2)=0.12$~MeV with $a=-18.5$~fm.
In this regard,
the Zeeman shift $h\lesssim 10$~MeV is not necessarily negligible in the dilute region.
Based on the theoretical study of spin-imbalanced  Fermi gases~\cite{PhysRevA.74.063628},
the saturation Zeeman shift $h_{\rm s}$ beyond which neutrons are fully polarized is expressed in terms of the attractive Fermi polaron energy $E_{\rm P}<0$ as (see also Fig.~\ref{fig:2}(a))
\begin{align}
    h_{\rm s}=\frac{E_{\rm F}+|E_{\rm P}|}{2}.
\end{align}
Note that $E_{\rm P}$ has been determined precisely in strongly interacting ultracold Fermi gases~\cite{PhysRevLett.102.230402,PhysRevLett.118.083602} (for review, see e.g.,~\cite{massignan2014polarons,tajima2021polaron,scazza2022repulsive}).
For a given density $\rho_{+1/2}$ of spin $s_z=+1/2$ neutrons where the direction of $B$ is parallel with $s_z=-1/2$,
the $G$-matrix calculation with the contact-type interaction reads~\cite{PhysRevA.84.033607,PhysRevA.107.053313}
\begin{align}
    E_{\rm P}&=\frac{\rho_{+1/2}}{\frac{M}{4\pi a}-\frac{Mk_{\rm F}}{2\pi^2}}
    =-\frac{2}{3}E_{\rm F}\frac{1}{1-\frac{\pi}{2}(k_{\rm F}a)^{-1}},
\end{align}
which gives $E_{\rm P}\simeq -0.67E_{\rm F}$ at unitarity (i.e., $(k_{\rm F}a)^{-1}=0$).
Regardless of its simple calculation, this value is \red{consistent with} the experimental value $E_{\rm P}=-0.64(7)E_{\rm F}$~\cite{PhysRevLett.102.230402}.
In the dilute limit, one obtains $E_{\rm P}/E_{\rm F}=\frac{4\pi a}{M}\rho_{\uparrow}/E_{\rm F}=\frac{4}{3\pi}(k_{\rm F}a)$, reproducing the Hartree shift.
We note that the finite-range effect in the nucleon-nucleon interaction can enlarge $E_{\rm P}$ as $E_{\rm P}\simeq 0.75E_{\rm F}$ at $k_{\rm F}\simeq 1$~fm$^{-1}$~\cite{PhysRevC.103.L052801}.
A similar tendency can also be found in the case of polaronic protons in neutron matter~\cite{tajima2023polaronic}.
Eventually, the saturation magnetic field reads
\begin{align}
\label{eq:bs}
    B_{\rm s}=
    \frac{E_{\rm F}}{|\gamma_n|}
    \left[1+\frac{2}{3}\frac{1}{1-\frac{\pi}{2}(k_{\rm F}a)^{-1}}\right].
\end{align}
Also, the Chandrasekhar-Clogston limit reads
$2h_{\rm c}/\mu_{\rm c}\simeq 1.09$ 
~\cite{frank2018universal,PhysRevA.103.043330},
where $h_{\rm c}$ and $\mu_{\rm c}$ are the critical values of the Zeeman shift and the chemical potential, respectively.
For simplicity, we take $\mu_{\rm c}\simeq \xi_{\rm B} E_{\rm F}$ where $\xi_{\rm B}\simeq 0.37$ is the Bertsch parameter~\cite{ku2012revealing,navon2010equation,PhysRevX.7.041004}.
Resulting $h_{\rm c}\simeq 0.2E_{\rm F}\equiv |\gamma_{\rm n}B_{\rm c}|/2$ gives a reasonable estimation around unitarity compared to the diagrammatic approach~\cite{PhysRevB.107.054505}.
Because of anisotropic scattering in polarized matter, the Fermi sea may be deformed.
Even in such a case, the Fermi polaron energy can be calculated as in Ref.~\cite{PhysRevA.103.033324}.

Figure~\ref{fig:2}(b) shows the ground-state phase diagram of dilute neutron matter with respect to $k_{\rm F}$ and $B$.
Above $B=B_{\rm s}$, the system is fully polarized and therefore the $^3P_0$ neutron superfluid can be a possible ground state in this region.
On the other hand, below $B=B_{\rm c}$
the system is expected to be well-studied $^1S_0$ neutron superfluid phase.
Suppose that the possible magnetic field reach $B\sim 10^{15\sim 18}$~G inside of the compact stars~\cite{broderick2000equation},
one may expect the spin-polarized $^3P_0$ neutron superfluid phase in the dilute region at $k_{\rm F}\lesssim 1$~fm$^{-1}$.
For comparison, the mean-field result with the separable interaction for the unpaired magnetic field \red{at $T=0.05$~MeV}~\cite{PhysRevC.93.015802} is plotted. On the one hand, it is larger than $B_{\rm c}$ at low densities. 
\red{This may originate from the difference of theoretical frameworks, where we have estimated $B_{\rm c}$ by using the results of cold-atom experiments, while the mean-field approximation was employed in Ref.~\cite{PhysRevC.93.015802}.
Indeed, the mean-field calculation overestimates the critical magnetic field in a strongly interacting regime~\cite{PhysRevA.90.053629}. On the other hand, the result of Ref.~\cite{PhysRevC.93.015802} drops to zero around $k_{\rm F}=0.8$~fm$^{-1}$, indicating the importance of the finite-range properties of the interaction.}
\redd{Because our approach is based on cold atomic physics where the finite-range effect is negligible,
our result is quantitatively valid in the low-density region $k_{\rm F}\ll r_{\rm eff}^{-1}\simeq 0.36$~fm$^{-1}$ (where $r_{\rm eff}=2.8$ fm is the $^1S_0$ effective range~\cite{PhysRevC.51.38}). In this regard, we used the dashed lines at $k_{\rm F}\gesim 0.1$ fm$^{-1}$ in Fig.~\ref{fig:2}(b).}
In the region between $B_{\rm s}$ and $B_{\rm c}$,
the ground state picture is elusive as various possibilities such as Sarma phase~\cite{sarma1963influence}, Fulde-Ferrel-Larkin-Ovchinikov superfluid~\cite{PhysRev.135.A550,larkin1965nonuniform}, induced $P$-wave pairing~\cite{PhysRevLett.97.020402}, and spin polarized droplet~\cite{PhysRevA.100.033613,PhysRevA.104.033304}
were discussed
(for a review see Refs~\cite{radzihovsky2010imbalanced,gubbels2013imbalanced}).
Since we are interested in the fully spin-polarized phase, we do not go details about this region.
However, if there exists the spin-polarized component, 
$^3P_0$ superfluid may appear in such a regime.
We note that the $^3P_2$ superfluid phase is not shown in Fig.~\ref{fig:2}(b) because it is expected to be found at larger densities~\red{\cite{RevModPhys.75.607,PhysRevC.104.045803}}.

\noindent{\it $^3P_0$ neutron superfluid theory}.---
We consider a neutron matter with spin-triplet $^3P_0$ interaction where the Hamiltonian is given by
    $H=K+V_{^3P_0}$.
The kinetic term $K$ of neutrons with the spin $s_z=\pm 1/2$, the mass $M=939$~MeV, and the chemical potential $\mu$ reads
\begin{align}
\label{eq:K}
    K&=\sum_{\bm{k}}\sum_{s_z=\pm 1/2}
    \left(\frac{k^2}{2M}-\mu-2s_z h\right)c_{\bm{k},s_z}^\dag
    c_{\bm{k},s_z}\cr
    &\simeq 
    \sum_{\bm{k}}
    \left(\frac{k^2}{2M}-\mu-h \right)c_{\bm{k},+1/2}^\dag
    c_{\bm{k},+1/2}.
\end{align}
In Eq.~\eqref{eq:K}, we ignored the $s_z=-1/2$ component by assuming the large Zeeman shift $h=|\gamma_{\rm n}B|/2>h_{\rm s}$ where the direction of $B$ is taken to be anti-parallel with respect to $s_z=+1/2$. 

For convenience, we define the $z$-projection of the two-neutron pair spin $S_z=s_z+s_z'$ and the total angular momentum $J_z=S_z+m$.
The $^3P_0$ interaction ($S=1$, $\ell=1$, $J=0$)  is given by
\begin{align}
    V_{^3P_0}
    &=2\pi\sum_{\bm{k},\bm{k}',\bm{P}}
    \sum_{m}
    \sum_{S_z}
    \sum_{s_z,s_z'}
    V(k,k')
    Y_{1,m}(\hat{\bm{k}})Y_{1,m}^*(\hat{\bm{k}}')\cr
    &\times\langle 
    1,m;1,S_z|\red{0},J_z\rangle^2
    \langle s,s_z;s,s_z|1,S_z\rangle^2
    \cr
    &\times 
    c_{\bm{k}+\bm{P}/2,s_z}^\dag
    c_{-\bm{k}+\bm{P}/2,s_z'}^\dag
    c_{-\bm{k}'+\bm{P}/2,s_z'}
    c_{\bm{k}'+\bm{P}/2,s_z},
\end{align}
where $\langle\ell,m;S,S_z|J,J_z\rangle$ is the Clebsh-Gordan coeffcient. 
For the spin-polarized dilute neutron matter,
we consider the $^3P_0$ interaction between two neutrons with the parallel spins (i.e., $s_z=s_{z'}=+1/2$) at zero center-of-mass momentum ($\bm{P}=\bm{0}$)
as
\begin{align}
    V_{^3P_0}&\simeq 2\pi\sum_{\bm{k},\bm{k}'}
    V(k,k')
    \left[
    \frac{1}{3}
    Y_{1,-1}(\hat{\bm{k}})Y_{1,-1}^*(\hat{\bm{k}}')
    \right]\cr
    &\times 
    c_{\bm{k},1/2}^\dag
    c_{-\bm{k},1/2}^\dag
    c_{-\bm{k}',1/2}
    c_{\bm{k}',1/2},
\end{align}
Furthermore, we use the separable interaction $V(k,k')=g\gamma_k\gamma_{k'}$ ($g<0$).
The superfluid order parameter is introduced as
\begin{align}
\label{eq:delta}
    \Delta_{\bm{k}}
    &= \gamma_k\frac{k_x-ik_y}{\sqrt{2}k}d,
\end{align}
where 
\begin{align}
\label{eq:d}
    d=-g\sum_{\bm{q}}\frac{q_x+iq_y}{\sqrt{2}q}\gamma_{q}
    \langle c_{-\bm{q},1/2}c_{\bm{q},1/2}\rangle.
\end{align}
Using Eqs.~\eqref{eq:delta} and \eqref{eq:d}, we obtain the mean-field Hamiltonian
\begin{align}
    H_{\rm MF}
    &=
    \frac{1}{2}
    \sum_{\bm{k}}
    \Psi_{\bm{k}}^\dag
    \left(\begin{array}{cc}
       \xi_{\bm{k},+1/2}  & -\Delta_{\bm{k}} \\
       -\Delta_{\bm{k}}^*  & -\xi_{\bm{k},+1/2}
    \end{array}\right)
    \Psi_{\bm{k}}\cr
    &\quad -\frac{|d|^2}{2g^2}
    +\frac{1}{2}\sum_{\bm{k}}\xi_{\bm{k},+1/2},
\end{align}
where $\Psi_{\bm{k}}=( c_{\bm{k},+1/2} \ c_{-\bm{k},+1/2}^\dag )^{\rm T}$ is the two-component Nambu spinor and $\xi_{\bm{k},+1/2}=k^2/2M-\mu-h$.
Accordingly, the Nambu-Gorkov Green's function is given by
$\hat{G}(\bm{k},i\omega_n)=\left[i\omega_n-\xi_{\bm{k},+1/2}\tau_3+{\rm Re}(\Delta_{\bm{k}})\tau_1-{\rm Im}(\Delta_{\bm{k}})\tau_2\right]^{-1}$
where $\tau_{1,2,3}$ is the Pauli matrix,
$\omega_n=(2n+1)\pi T$ is the fermion Matsubara frequency ($n\in \mathbb{Z}$), and
$E_{\bm{k}}=\sqrt{\xi_{\bm{k},+1/2}^2+|\Delta_{\bm{k}}|^2}
$ is the BCS quasiparticle dispersion.
The gap equation is obtained from
$d=gT\sum_{\bm{q}}\sum_{i\omega_n}\frac{q_x+iq_y}{\sqrt{2}q}\gamma_q G_{12}(\bm{q},i\omega_n)$, which can be rewritten as
\begin{align}
\label{eq:gapeq}
    1=-g\sum_{\bm{q}}\frac{q_x^2+q_y^2}{2q^2}\frac{\gamma_q^2}{2E_{\bm{q}}}\tanh\left(\frac{E_{\bm{q}}}{2T}\right).
\end{align}
By solving Eq.~(\ref{eq:gapeq}), one can examine the $^3P_0$ superfluid properties in spin-polarized neutron matter.

We note that the present $^3P_0$ neutron superfluid with momentum-dependent gap $\Delta_{\bm{k}}\propto k_x-ik_y$ ($\Delta_{\bm{k}}^*\propto k_x+ik_y$) is similar to the gapless $p_x+ip_y$ Fermi superfluid predicted in cold atoms~\cite{PhysRevLett.94.230403} as well as the $A_1$ phase in $^3$He superfluids~\cite{PhysRevA.8.1633,PhysRevLett.30.81,PhysRevLett.33.686}.
Indeed, the gapless points (called the Weyl nodes or the Fermi points) can be found at $E_{\bm{k}}=0$, that is, $\bm{k}=(0,0,\pm\sqrt{2M(\mu+h)})$ in the momentum space~\cite{volovik2007quantum}.
Thus, if the $^3P_0$ superfluid is present, the superfluid spin edge current may flow accompanying the chiral anomaly around the surface of uniform neutron matter.
Moreover, the $^3P_0$ gap structure leads to an anisotropic spin response, which may induce a characteristic spin transport at the interfaces with other phases (e.g., $^1S_0$ superfluid) as in the case of superconducting materials~\cite{PhysRevLett.127.207001,PhysRevB.106.L161406,PhysRevB.107.184437}. 
Another important consequence of the $^3P_0$ superfluid can be found in the specific heat $C$, which plays a crucial role in the thermal transport during the cooling process~\cite{PhysRevC.87.035803}.
Because of the gapless quasiparticle excitation, $C\propto T^3$ can be found at low temperature~\cite{kagan2010bcs}.
Also, the pair-breaking formation process associated with $^3P_0$ superfluid might occur in the cooling process~\cite{leinson2015superfluid}.
The topological aspects of $^3P_0$ superfluid and associated vortices can be further examined as in the case of $^3P_2$ superfluid in the core region~\cite{PhysRevC.104.045803,PhysRevC.105.035807,PhysRevB.105.L220503}.

Next, we relate the model parameters with the $^3P_0$ scattering amplitude
    $f_{^3P_0}(k)=k^2\left(-v^{-1}+\frac{1}{2}rk^2-ik^3\right)^{-1}$,
where $v$ and $r$ are the scattering volume and the effective range, respectively. 
These low-energy constants are given by 
$v=-2.638$ fm$^{3}$ and $r=3.182 \ {\rm fm}^{-1}$
~\cite{PhysRevC.82.034003,PhysRevC.29.739}.
The two-body $T$-matrix reads~\cite{gurarie2007resonantly}
\begin{align}
    T(\bm{p},\bm{p}';\omega)=\bar{V}(\bm{p},\bm{p}')+\sum_{\bm{q}}\frac{\bar{V}(\bm{p},\bm{q})T(\bm{q},\bm{p}';\omega)}{\omega_+-q^2/M_\nu},
\end{align}
with $\omega_+=\omega+i\eta$ ($\eta$ is an infinitesimally small value) and
$\bar{V}(\bm{p},\bm{p}')=\frac{4\pi}{3}g\gamma_{p}\gamma_{p'}Y_{1,-1}(\hat{\bm{p}})Y_{1,-1}^*(\hat{\bm{p}}')$. 
The separability of the $T$-matrix leads to the form given by
   $ T(\bm{p},\bm{p}';\omega)=\frac{4\pi}{3}t(\omega_+)\gamma_{p}\gamma_{p'}Y_{1,-1}(\hat{\bm{p}})Y_{1,-1}^*(\hat{\bm{p}}')$,
where
    $t(\omega)
    =g\left[1-g\Pi(\omega_+)\right]^{-1}$.
The pair propagator $\Pi(\omega)$ is given by
\begin{align}
    \Pi(\omega)&=\sum_{\bm{q}}\frac{4\pi}{3}\gamma_{q}^2
    Y_{1,-1}^*(\hat{\bm{q}})Y_{1,-1}(\hat{\bm{q}})\frac{1}{\omega_+-q^2/M}.
\end{align}
For simplicity, we employ
    $\gamma_{q}=\frac{q}{1+(q/\Lambda)^2}$.
In this case, we obtain
\begin{align}
    \Pi(\omega)
    &=-\frac{iM\Lambda^4(2\sqrt{M\omega_+}+i\Lambda)(\sqrt{M\omega}-i\Lambda)^2}{24\pi(M\omega+\Lambda^2)^2}.
\end{align}
Using this, we find the scattering amplitude
    $f_{^3P_0}(k)
    =
    k^2\left[\frac{12\pi(k^2+\Lambda^2)^2}{M g\Lambda^4}+\frac{i(2k+i\Lambda)(k-i\Lambda)^2}{2}\right]^{-1}$.
In this way, we find the condition for $g$ and $\Lambda$
$
    v^{-1}=\frac{12\pi}{M}\left(\frac{1}{g}+\frac{M\Lambda^3}{24\pi}\right)$,
$    r=-\frac{24\pi}{M}\left(\frac{2}{g\Lambda^2}+\frac{M\Lambda}{8\pi}\right)=-\frac{48\pi}{gM\Lambda^2}-3\Lambda$.
For $v=-2.638$ fm$^{-3}$ and $r=3.182$ fm$^{-1}$,
we obtain $\Lambda=0.63058$ fm$^{-1}$ and $Mg=-74.745$~fm$^{3}$.

\noindent{\it Critical temperature of $^3P_0$ superfluid}.---
The superfluid critical temperature $T_{\rm c}$ is an important quantity to examine the possible appearance of the $^3P_0$ neutron superfluid in an astrophysical environment.
We can calculate $T_{\rm c}$ by taking $T\rightarrow T_{\rm c}$ and $d\rightarrow 0$ in Eq.~\eqref{eq:gapeq} as
\begin{align}
\label{eq:tc}
1&=-\frac{Mg}{6\pi^2}
\int_0^{\infty}q^2dq
\frac{\gamma_q^2}{2M \xi_{\bm{q},+1/2}}{\rm tanh}\left(\frac{\xi_{\bm{q},+1/2}}{2T_{\rm c}}\right).
\end{align}
Eq.~\eqref{eq:tc} should be solved together with the number density equation~\cite{leggett2008diatomic}
\begin{align}
    \rho_{+1/2}
    &=\frac{1}{4\pi^2}
    \int_0^{\infty}k^2dk
    \left[1-{\rm tanh}\left(\frac{\xi_{\bm{k},+1/2}}{2T_{\rm c}}\right)\right].
\end{align}

\begin{figure}[t]
    \centering
    \includegraphics[width=8.6cm]{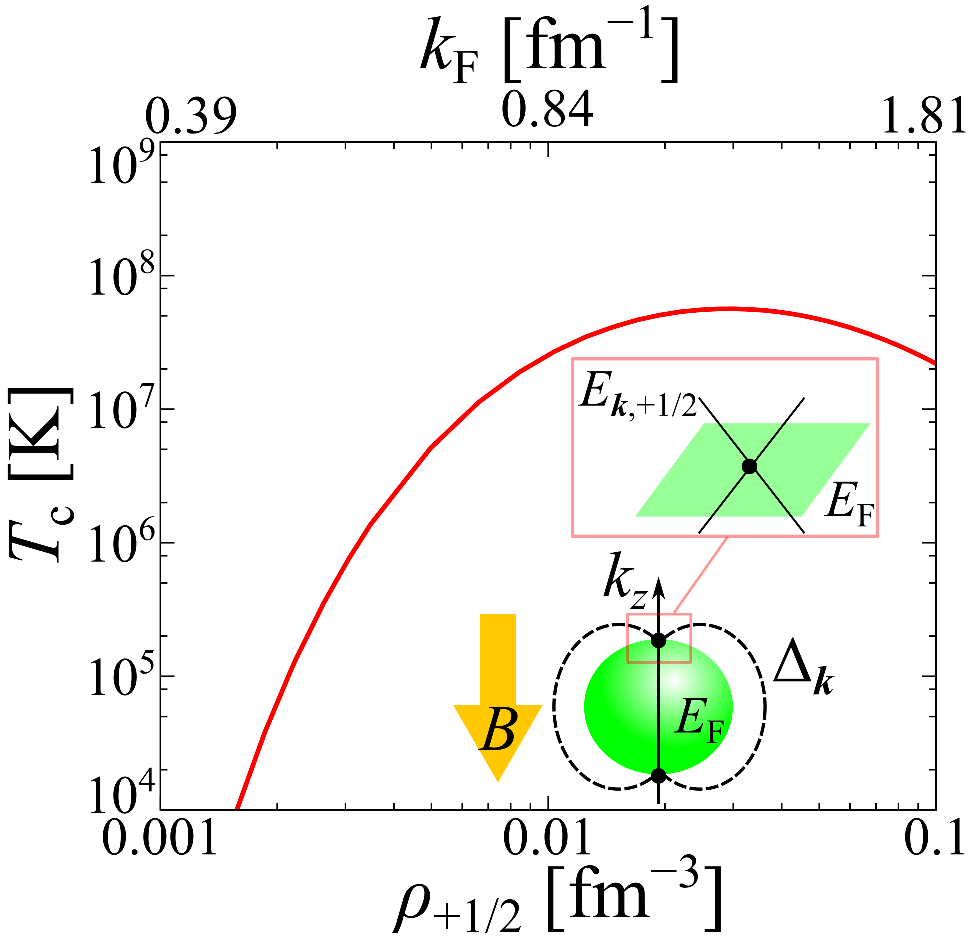}
    \caption{Critical temperature $T_{\rm c}$ of $^3P_0$ neutron superfluidity exhibiting the Weyl nodes at $\bm{k}=(0,0,\pm\sqrt{2M(\mu+h)})$ as a function of the spin $s_z=+1/2$ neutron density $\rho_{+1/2}$.
    \red{For reference, we show the corresponding neutron Fermi momenta $k_{\rm F}=(6\pi^2\rho_{+1/2})^{1/3}$ above.}
    }
    \label{fig:3}
\end{figure}

Figure~\ref{fig:3} shows the calculated critical temperature $T_{\rm c}$ of $^3P_0$ neutron superfluidity as a function of $\rho_{+1/2}$.
$T_{\rm c}$ reaches $10^7$~K around $\rho_{+1/2}\simeq0.01$~fm$^{-3}$.
This temperature can be realized in old neutron stars~\cite{prakash2001evolution}.
Although the temperature in a magnetar is typically larger than those of old stars~\cite{PhysRevLett.102.091101}, $T\sim 10^7$~K may be not so unrealistic.
In this regard, we need to carefully consider the magnetic field and the temperature in neutron stars to explore the possible candidates of stars involving $^3P_0$ neutron superfluidity.
The suitable condition for $^3P_0$ superfluid would be $B>B_{\rm s}$ and $T<T_{\rm c}$.
\redd{It should be noted that the density region in Fig.~\ref{fig:3} is beyond the regime where the zero-range approximation for the $^1S_0$ channel in Fig.~\ref{fig:2} is quantitatively valid (i.e., $k_{\rm F}\ll 0.36$ fm$^{-1}$). In this sense, further quantitative investigation of $B_{\rm s}$ with the finite-range $^1S_0$ interaction is needed.
Nevertheless, $T_{\rm c}$ in Fig.~\ref{fig:3} would be useful once spin-polarized neutron matter is realized.}
On the other hand, it may be not so straightforward to precisely determine the profiles of $T$ and $B$ inside stars~\cite{PhysRevC.99.055811}. 
\red{As three-dimensional evolution of the magnetic field has been studied by the magneto-hydrodynamic simulations recently~\cite{10.1093/mnras/stad1773},
the survey of the regime satisfying this condition by combining the recent observations and simulations would be left as an interesting future work.}

\red{We note that while we employ the mean-field approximation to calculate $T_{\rm c}$ of $^3P_0$ superfluid, this calculation can be justified in the $^3P_0$ channel because the $^3P_0$ interaction strength itself is not so strong as we can see its phase shift in Fig.~\ref{fig:1}, which is much smaller than the $^1S_0$ one. Moreover, in the dilute region, the density and spin fluctuations are not crucial compared to those in the dense region as found in the $^1S_0$ channel~\cite{ramanan2021pairing}.}


\noindent{\it Summary}.---
In this work,
we have theoretically discussed the possible appearance of $^3P_0$ neutron superfluid, which was overlooked in the literature, in dilute neutron matter with the large spin polarization under the strong magnetic field.
For the ground state at $T=0$, we qualitatively clarified the spin-polarized $^3P_0$ superfluid phase in the phase diagram with respect to the magnetic field $B$ and the Fermi momentum $k_{\rm F}$ in the dilute regime.
After developing the superfluid theory for such a phase with the separable interaction,
we have numerically calculated the $^3P_0$ superfluid critical temperature as a function of the spin-polarized neutron density.
Our result would be useful for further understanding of possible exotic state of matter in neutron stars.

For further quantitative investigations, it is important to study the critical magnetic field of $^1S_0$ superfluid within the beyond-mean-field theory as well as the model with finite-range interactions.
One can also expect a $^3P_0$ pairing of protons which are dilute even around the core region of a neutron star. 
In such a case, the Landau quantization plays a crucial role~\cite{PhysRevC.75.034001}.
\redd{Moreover, while we have considered homogeneous neutron matter in this study, nuclear clusters would be important as the non-monotonic density dependence of the $S$-wave pairing gap has been reported in the presence of the inhomogeneous potential induced by nuclei~\cite{PhysRevLett.107.205301,PhysRevC.88.034314}.
A proximity effect~\cite{10.1093/ptep/ptaa174} associated with the topological $^3P_0$ superfluid can also be expected near the nuclear clusters.
Since neutrons inside the nuclei are insensitive to the magnetic field because of the large density compared to the gas phase,
the competition between $^1S_0$ and $^3P_0$ pairings may occur near the surface of neutron-rich nuclei in the crust.
}

\noindent
\textit{Note added}.---
When this paper was being finalized, there appeared a preprint~\cite{krotscheck2023triplet}, where the $^3P_0$ triplet superfluid in neutron matter is discussed.

\acknowledgements
H.~T. thank K.~Sekizawa, K.~Iida, Y.~Ominato, K. Yoshida, A. Dohi, and the members of H.~Liang group for the useful discussion and A.~Sedrakian for providing the numerical data in Ref.~\cite{PhysRevC.93.015802}.
N.~Y. also thank H.~Okawa and K.~Fujisawa for fruitful discussion on the evolutions of neutron stars.
The authors thank RIKEN iTHEMS NEW working group for fruitful discussions.
This work is supported by Grants-in-Aid for Scientific
Research provided by JSPS through Nos.~18H05406, 20K03951, 21H01800, 21H04565, 22H01158, 22K13981, and 23H01839.
Y.~S. is supported by Pioneering Program of RIKEN for Evolution of Matter in the Universe
(r-EMU).

\bibliographystyle{apsrev4-2}
\bibliography{reference.bib}



\end{document}